\documentclass[aps,amsmath,amssymb,a4paper,11pt,superscriptaddress]{revtex4-2}
\usepackage{graphicx}
\usepackage{dcolumn}
\usepackage{bm}

\usepackage{xspace}

\newcommand*{\Nch}{\ensuremath{N_\mathrm{ch}}\xspace}
\newcommand*{\Nfw}{\ensuremath{N_\mathrm{fw}}\xspace}
\newcommand*{\RT}{\ensuremath{R_\mathrm{T}}\xspace}
\newcommand*{\RNC}{\ensuremath{R_\mathrm{NC}}\xspace}
\newcommand*{\SO}{\ensuremath{S_\mathrm{0}}\xspace}
\newcommand*{\pT}{\ensuremath{p_\mathrm{T}}\xspace}

\newcommand*{\GeVc}{\ensuremath{\mathrm{GeV}/c}\xspace}

\newcommand*{\Lc}{\ensuremath{\Lambda_{\rm c}^+}\xspace}
\newcommand*{\Xic}{\ensuremath{\Xi_{\rm c}^{0,+}}\xspace}
\newcommand*{\Sigc}{\ensuremath{\Sigma_{\rm c}^{0,++}}\xspace}
\newcommand*{\Dz}{\ensuremath{\rm D^0}\xspace}

\newcommand*{\LcToDz}{\ensuremath{{\Lambda_{\rm c}^+}/{\rm D^0}}\xspace}

\begin{document}


\title[]{The role of the underlying event in the \Lc enhancement in high-energy pp collisions}

\author{Zolt\'{a}n Varga}
\altaffiliation{Correspondence: varga.zoltan@wigner.hu}
\affiliation{Wigner Research Centre for Physics, P.O. Box 49, H-1525 Budapest, Hungary}%
\affiliation{Department of Theoretical Physics, Budapest University of Technology and Economics, Budafoki \'ut 8, H-1111 Budapest, Hungary}%
\author{R\'{o}bert V\'ertesi}
\affiliation{Wigner Research Centre for Physics, P.O. Box 49, H-1525 Budapest, Hungary}%

\date{\today}

\begin{abstract}
We study the enhanced production of charmed baryons relative to charmed mesons in proton--proton collisions at LHC energies. Utilizing PYTHIA 8 simulations with enhanced color-reconnection we propose methods based on the comparative use of several event-activity classifiers to identify the source of the charmed-baryon enhancement. We also conclude that in the scenario under investigation the excess \Lc production is primarily linked to the underlying event activity and not to the production of jets.
\end{abstract}


\maketitle

\section{Introduction}
\label{sec:introduction}

The production of heavy-flavor hadrons in high-energy collisions is usually described using the factorization approach~\cite{Collins:1989gx}, in which the production cross section is expressed as convolutions of three independent factors: the parton distribution functions (PDF) of the colliding hadrons, the production cross-sections of the heavy quarks in the hard partonic process, and the fragmentation functions of the heavy quarks into the given heavy-flavor hadron species. Fragmentation has been assumed to be universal, independent of the collision system. 
The ALICE and CMS experiments at the LHC observed a low-momentum enhancement in the production of charmed \Lc baryons compared to charmed \Dz mesons in high-energy proton-proton collisions, compared to model calculations tuned for electron-positron colllisions~\cite{ALICE:2017thy,CMS:2019uws,ALICE:2020wfu}. Since the PDF and partonic cross sections cancel in the ratio, this suggests that the universality of charm fragmentation is not fulfilled. 
There are several recent works, however, that explain the excess with different scenarios, such as color reconnection (CR) beyond leading color approximation \cite{Christiansen:2015yqa}, quark recombination mechanism (QCM) \cite{Song:2018tpv} or coalescence \cite{Plumari:2017ntm}, and feed-down from several higher mass charm states \cite{He:2019tik}.
Although all these scenarios tend to qualitatively describe the trends observed for the \Lc production, most of them fall short in explaining the yields and ratios of further charmed baryonic states e.g. \Xic and \Sigc \cite{ALICE:2021bli}.
Preliminary data show that the \Lc enhancement correlates with event-multiplicity~\cite{Hills:2021eto}. Trends observed in the event multiplicity are reproduced well with models containing CR beyond leading color approximation. This is expected since in this latter scenario, CR is linked to multiple-parton interactions (MPI), that in turn is related to the activity of the underlying event (UE)~\cite{Martin:2016igp} and thus to the event multiplicity.

In the followings we study \Lc enhancement using simulations with color reconnection beyond leading color approximation, using event-activity classifiers that are sensitive to the origin of charm production. The methods we propose can be used in future measurements to achieve a high discriminatory power between the different scenarios. 

\section{Analysis technique}
\label{sec:simualtion}

We simulated 1 billion pp collisions at $\sqrt{s}=13$ $\mathrm{TeV}$ using the PYTHIA 8.303 Monte Carlo (MC) event generator~\cite{Sjostrand:2014zea} with soft QCD settings and the enhanced CR mode 2. Mode 2 is known to reproduce the trends in \LcToDz ratios well~\cite{ALICE:2020wfu,Hills:2021eto}. 
We  verified that simulations with the Monash tune without enhanced CR reproduce earlier simulation results~\cite{ALICE:2020wfu}, and also compared modes 0 and 3 to results with mode 2.
Mode 0 yields qualitatively similar results to mode 2 with slightly less enhancement, while mode 3 was found to vastly overestimate the underlying event. 

Particle tracks were selected at mid-rapidity, in the pseudorapidity window $|\eta|<1$ and in the full azimuth angle $\varphi$, with a minimum transverse momentum \pT$>0.15$~\GeVc.
The charmed \Lc and \Dz particles, as well as their charged conjugates, were selected in the rapidity window $|y|<0.5$ based on MC information. In the following, we refer to both the particles and their charge conjugates as \Lc and \Dz respectively. In case of cascade decays, only the last charm particle was considered in the chain.

To quantitatively characterise an event, first we used the event multiplicity \Nch, defined as the number of all charged final state particles in the event in the mid-rapidity acceptance defined above. We also used the forward multiplicity \Nfw, which we defined with the acceptance $2<|\eta|<5$ to ensure a rapidity gap and thus reduce autocorrelation.
In order to specifically characterize the event with the underlying-event activity, we utilized the 
transverse event-activity classifier $R_\mathrm{T} \equiv N_\mathrm{ch}^{\mathrm{transverse}}/\langle N_\mathrm{ch}^{\mathrm{transverse}}\rangle$~\cite{Martin:2016igp}, where the transverse region is defined in the azimuth angle with respect to a high-momentum trigger, $p_\mathrm{T}^{\mathrm{leading}}>5$~GeV/$c$ as $\frac{\pi}{3}<|\Delta \phi|<\frac{2\pi}{3}$. In models such as PYTHIA that describe events in terms of MPI, the \RT quantity is strongly correlated with the number of MPIs in an event~\cite{Martin:2016igp}.
Analogously, we defined the near-side cone activity $R_\mathrm{NC} \equiv N_\mathrm{ch}^{\mathrm{near\text{-}side\; cone}}/\langle N_\mathrm{ch}^{\mathrm{near\text{-}side\; cone}}\rangle$ in a narrow cone around the trigger particle, $\sqrt{\Delta\phi^2+\Delta\eta^2}<0.5$. As this region is dominated by the fragments of the jet containing the leading particle, $R_\mathrm{NC}$ will be primarily determined the multiplicity of the jet initiated by the leading hard process.

Finally we use the transverse spherocity variable~\cite{Ortiz:2015ttf} 
\begin{eqnarray}
\SO \equiv \frac{\pi}{4} \min\limits_{\bf \hat{n}} \left( \frac{\sum_i \left| {\bf p}_{{\rm T},i} \times {\bf \hat{n}} \right| }{\sum_i p_{{\rm T},i}} \right)\ , 
\end{eqnarray} 
where $i$ runs over all the particles in the acceptance and ${\bf \hat{n}}$ is any unit vector in the azimuth plane. Transverse spherocity characterizes the events by jettiness in the azimuthal plane between 0 and 1 by construction: for isotropic events \SO approaches unity, and for events determined by collimated prongs of particles, \SO is close to zero.
The event-activity-class limits, summarized in Table~\ref{tab:class}, were determined to contain roughly similar number of events.

\begin{table}[h]
\begin{center}
\begin{tabular}{ |c|c|c|c|c|c| } 
 \hline
 class & \#1 & \#2 & \#3 & \#4 & \#5 \\
 \hline\hline 
 \Nch & $\le$15 & 16--30 & 31--40 & 41--50 & $\ge$51 \\
 \Nfw & $\le$45 & 46--90 & 91--120 & 121--150 & $\ge$151 \\
 \RT & $<$0.5 & 0.5--1 & 1--1.5 & 1.5--2 & $>$2 \\
 \RNC & $<$0.5 & 0.5--1 & 1--1.5 & 1.5--2 & $>$2 \\
 \SO & 0--0.25 & 0.25--0.45 & 0.45--0.55 & 0.55--0.75 & 0.75--1 \\
 \hline
\end{tabular}
\caption{\label{tab:class}Definition of event classes with the different event-activity classifiers.}
\end{center}
\end{table}

\section{Results}
\label{sec:results}

\begin{figure}[ht!]%
\centering
\includegraphics[width=0.5\columnwidth]{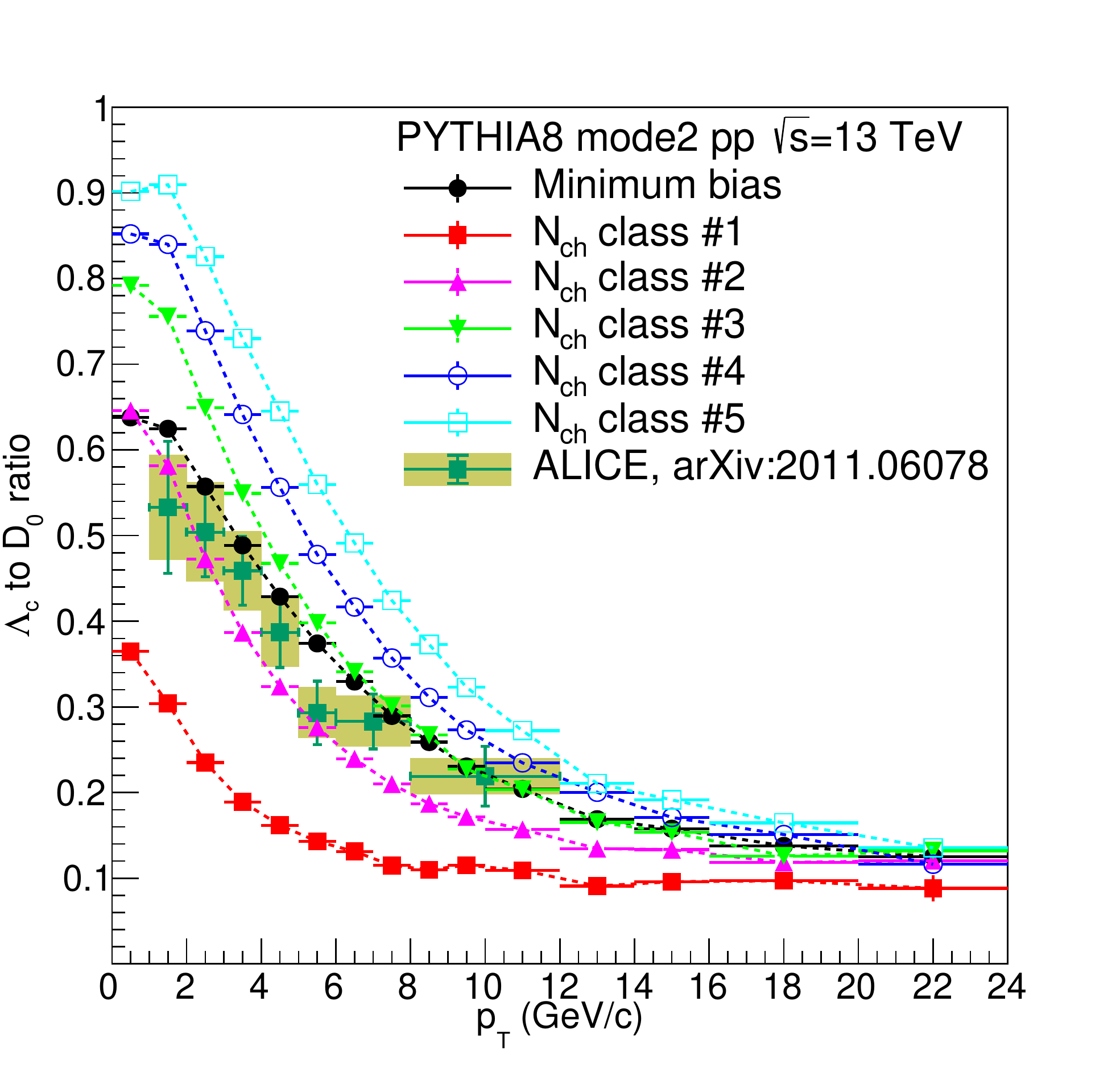}%
\includegraphics[width=0.5\columnwidth]{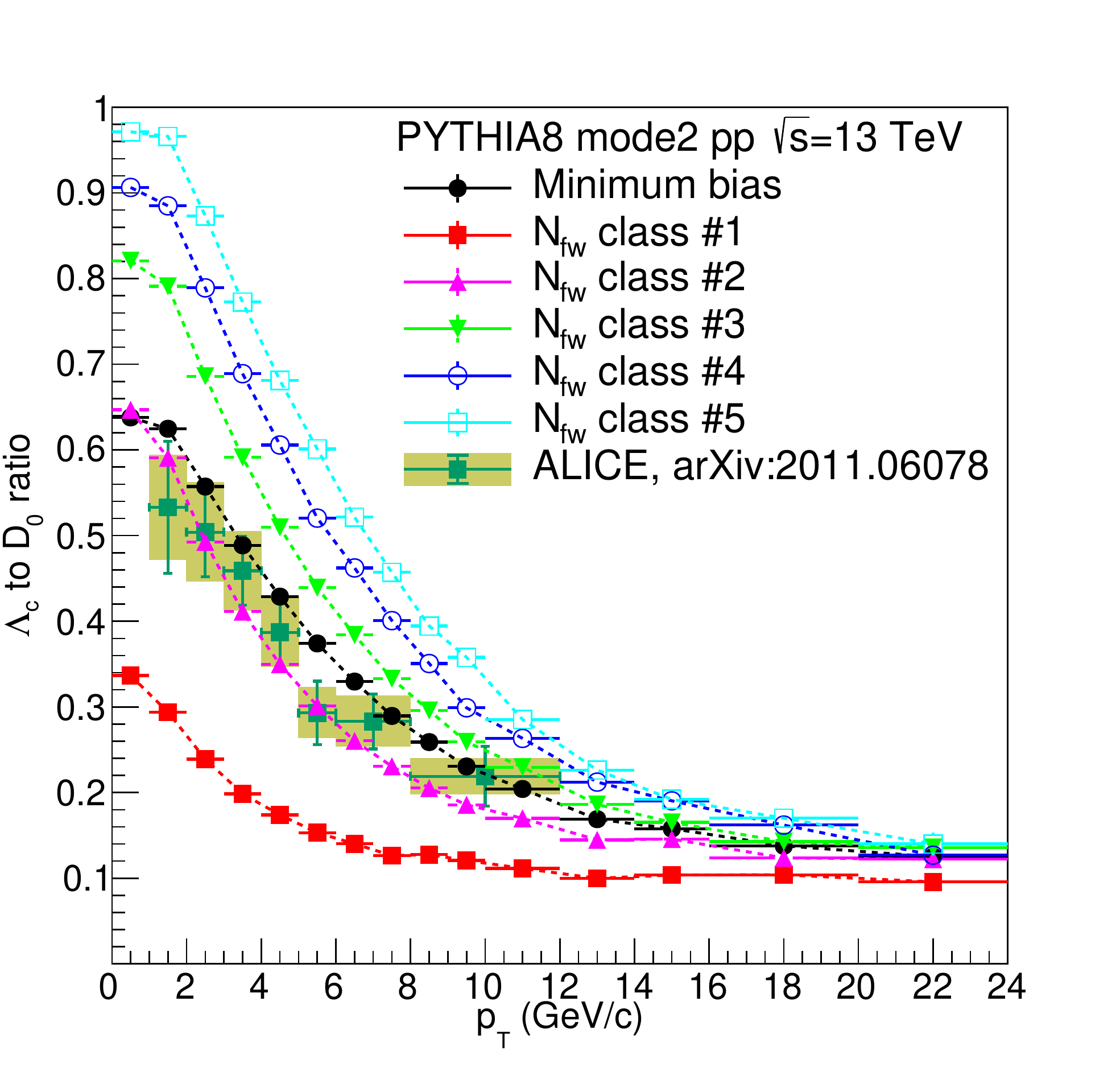}%
\caption{\Lc to \Dz ratios in function of \pT, for different mid-rapidity charged-hadron multiplicity bins (left) as well as forward multiplicity bins (right). Minimum-bias results are compared to ALICE measurements~\cite{ALICE:2020wfu} for reference.}
\label{fig:Mults}
\end{figure}

In Fig.~\ref{fig:Mults} (left) we compare the \Lc to \Dz ratios for different charged event multiplicity bins for CR mode 2. Minimum-bias results are compared to ALICE measurements~\cite{ALICE:2020wfu} for reference.
It can be seen that the \Lc yield has a significant dependence on the multiplicity, as it increases by increasing multiplicities. This qualitatively reproduces the behavior seen in~\cite{Hills:2021eto}. In Fig.~\ref{fig:Mults} (right) we observe a similar dependence on the forward multiplicity, the \Lc yield again increases for higher forward multiplicities.
In case of events classified by the forward multiplicity, a rapidity gap is present that reduces the correlation of the measured charmed-hadron yields from leading hard processes with the high charged-hadron multiplicity stemming from the same charmed jets. Since the patterns are similar, the multiplicity-dependence is not driven by charm-production in jets. 

\begin{figure}[ht!]%
\centering
\includegraphics[width=0.5\columnwidth]{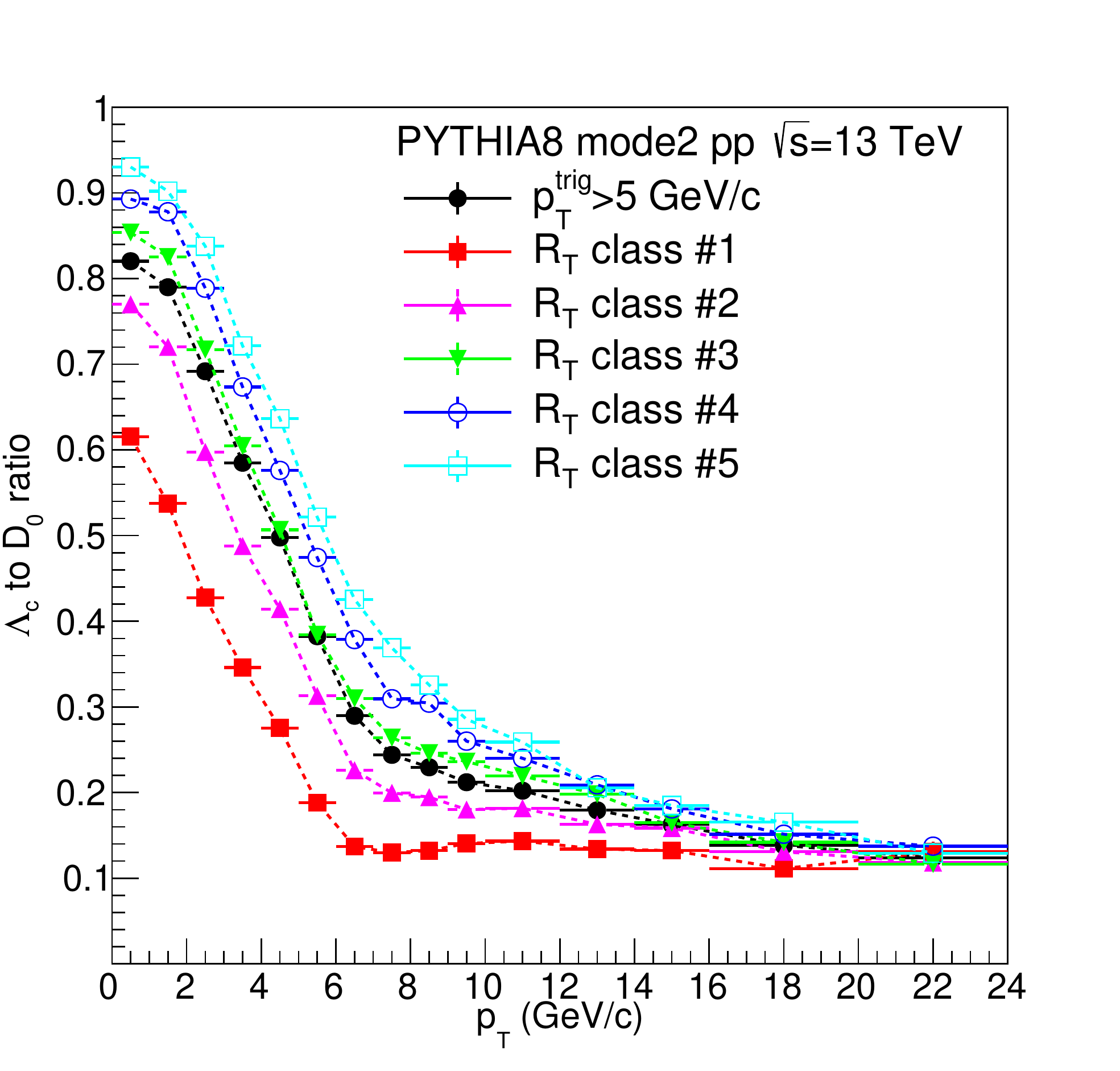}%
\includegraphics[width=0.5\columnwidth]{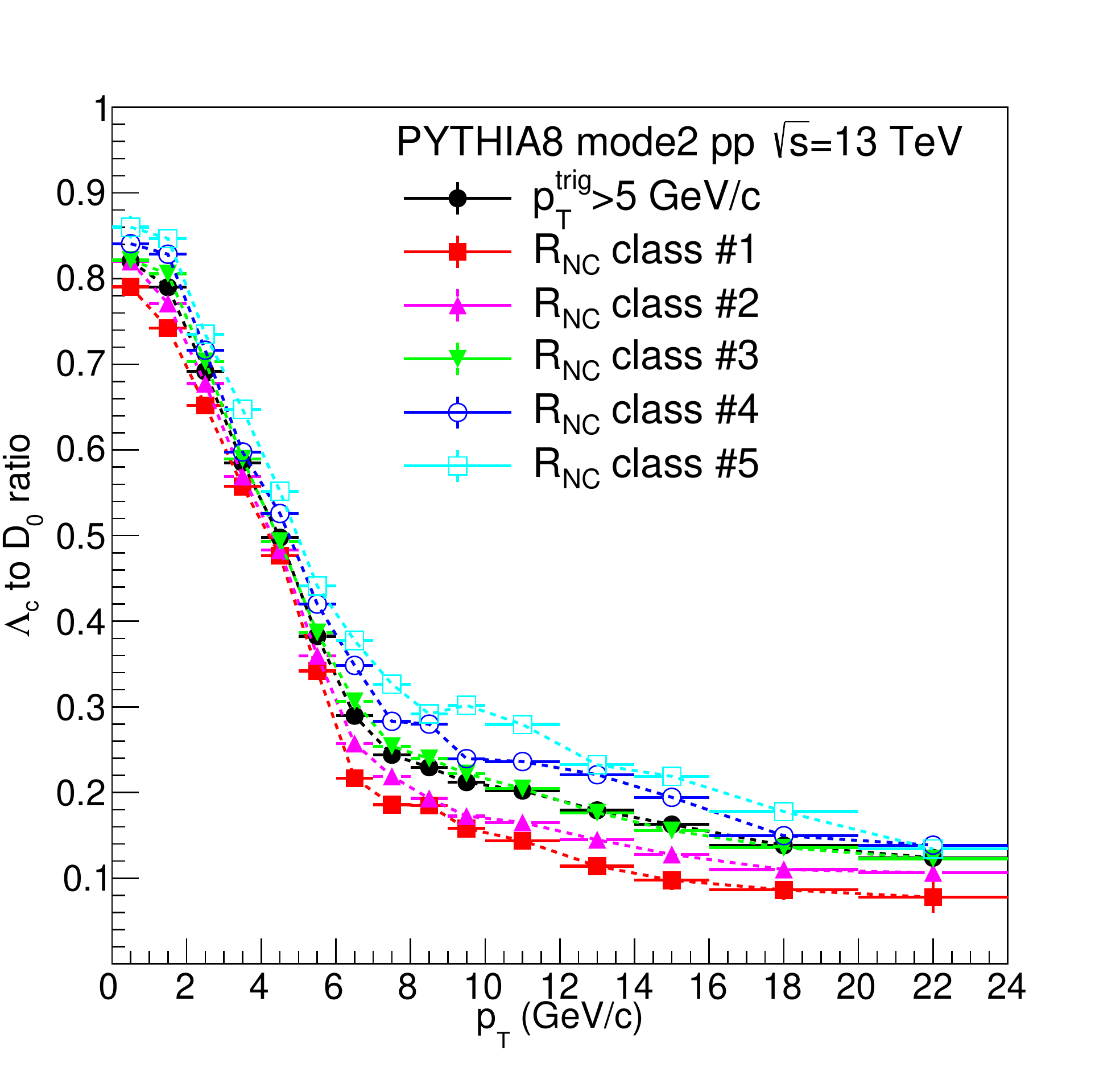}
\includegraphics[width=0.5\columnwidth]{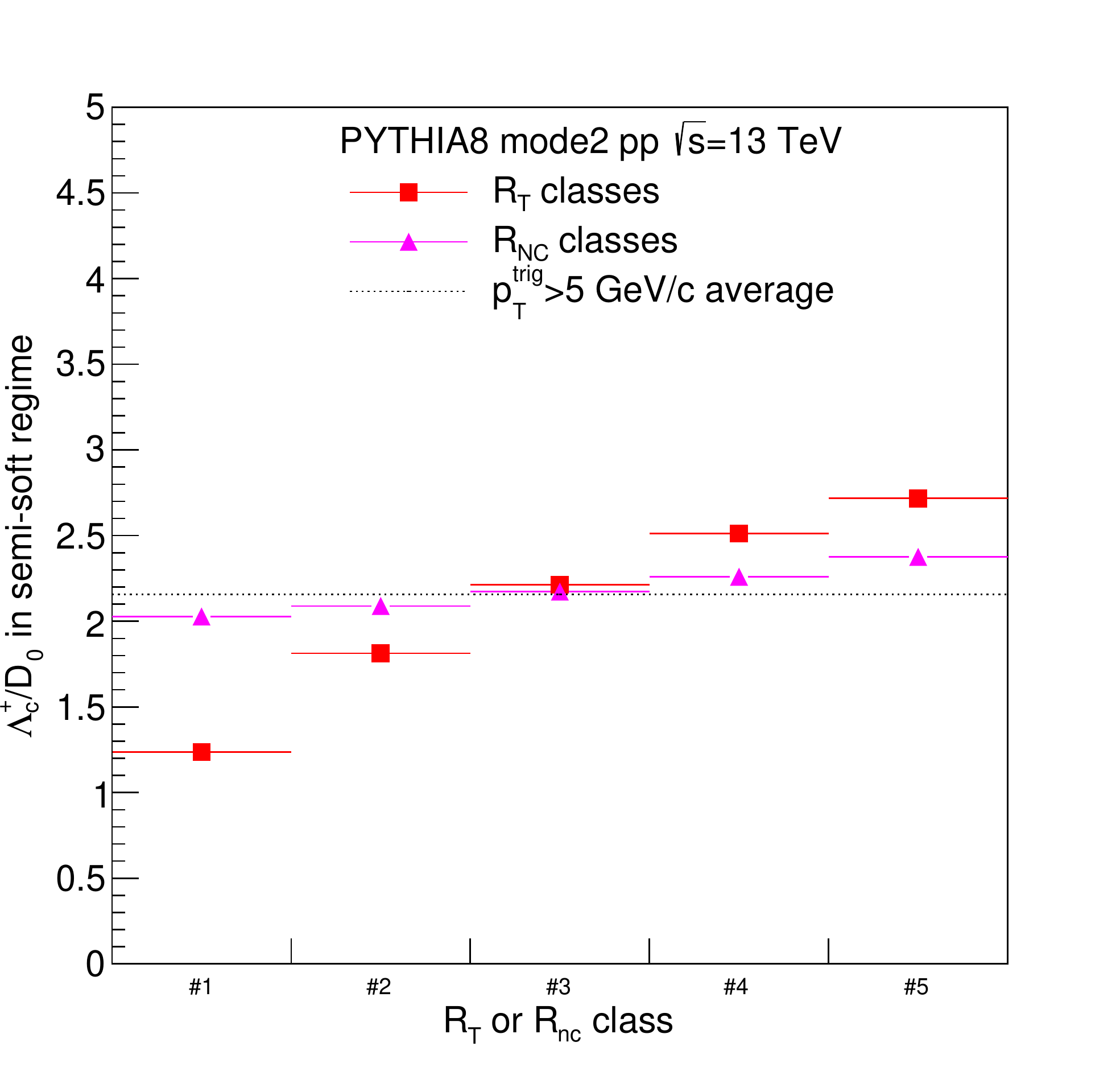}%
\caption{Top: \Lc to \Dz ratios of hadron-triggered events in function of \pT, for different \RT bins (top left) and \RNC bins (top right), together with multiplicity-inclusive hadron-triggered data, shown for reference. Bottom: \Lc to \Dz ratios integrated over the coalescence regime $2<\pT<6$ \GeVc in function of \RT and \RNC bins for hadron-triggered data. The dashed line represents the average of triggered events.}
\label{fig:RT_RNC}
\end{figure}

In the upper row of Fig.~\ref{fig:RT_RNC} we plot the \Lc to \Dz ratios as a function of \pT for different \RT and \RNC classes. 
In the case of the transverse-side multiplicities, higher \RT values correspond to stronger \Lc enhancement. On the other hand, for different near-side-cone multiplicities the \Lc to \Dz ratios remain consistent within fluctuations in the $\pT<6$ GeV/$c$ range, while there is some difference for $\pT>6$ GeV/$c$. We can conclude therefore, that the increased \Lc yield is primarily connected to charm production within the underlying event and not the jet region.
This is further highlighted in the bottom panel of Fig.~\ref{fig:RT_RNC}, where we show the \Lc to \Dz ratio integrated over the semi-soft (or coalescence) regime $2 <\pT<6$ GeV/$c$, plotted in function of \RT as well as \RNC. While the change from small to large \RT values is almost threefold, there is a very slight dependence on \RNC, which may be caused by the UE that is not subtracted from the jet region.

A shortcoming of event classification based on \RT and \RNC is that it requires a high-\pT trigger, thus it restricts the analysis to events containing a hard process. This introduces a bias into the sample and makes it more difficult to accumulate sufficient statistics in an experimental environment. The "jettiness" of an event, however, can be defined with the help of \SO regardless of the presence of a hard process. 

\begin{figure}[ht!]%
\centering
\includegraphics[width=0.5\columnwidth]{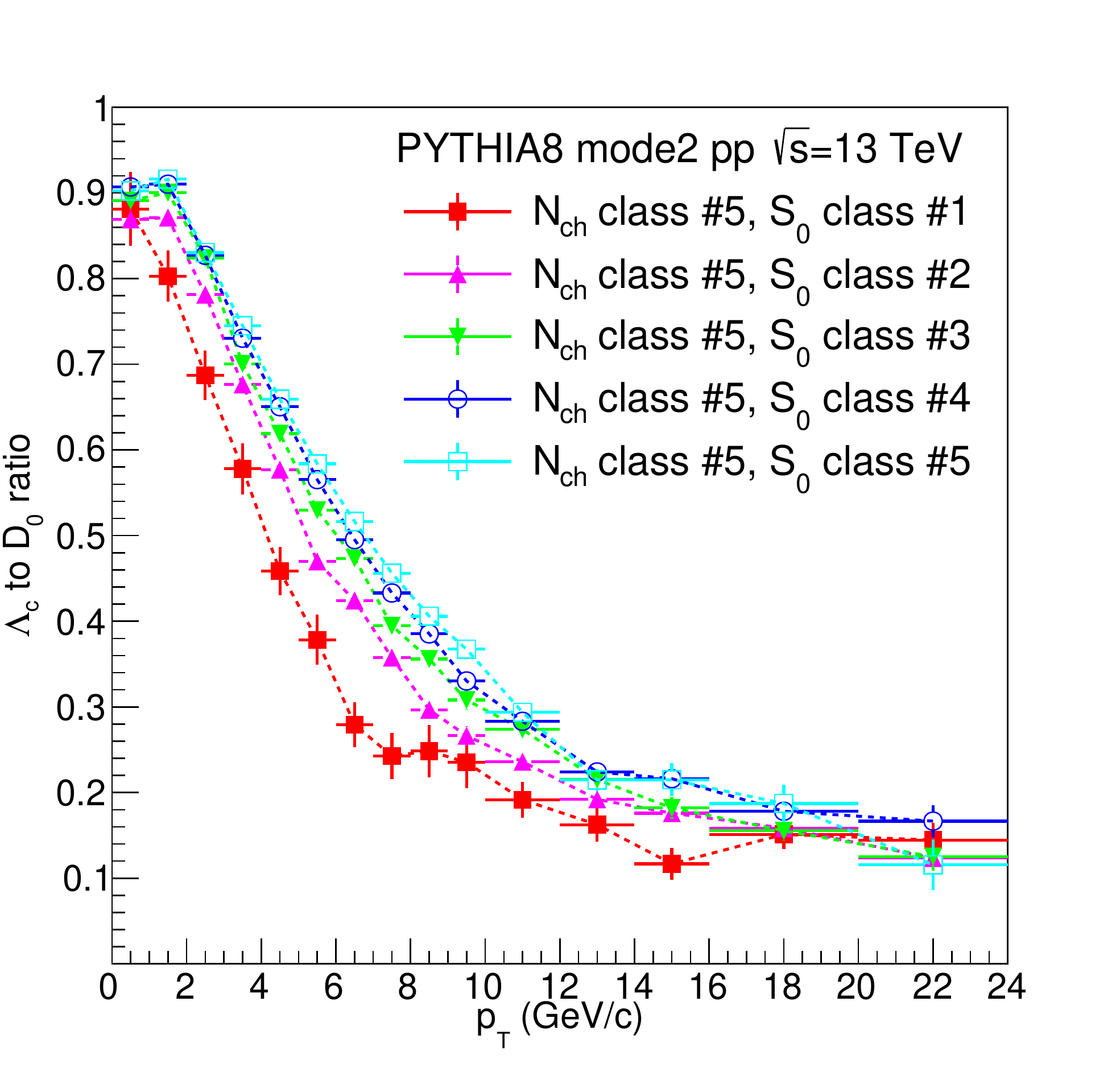}%
\includegraphics[width=0.5\columnwidth]{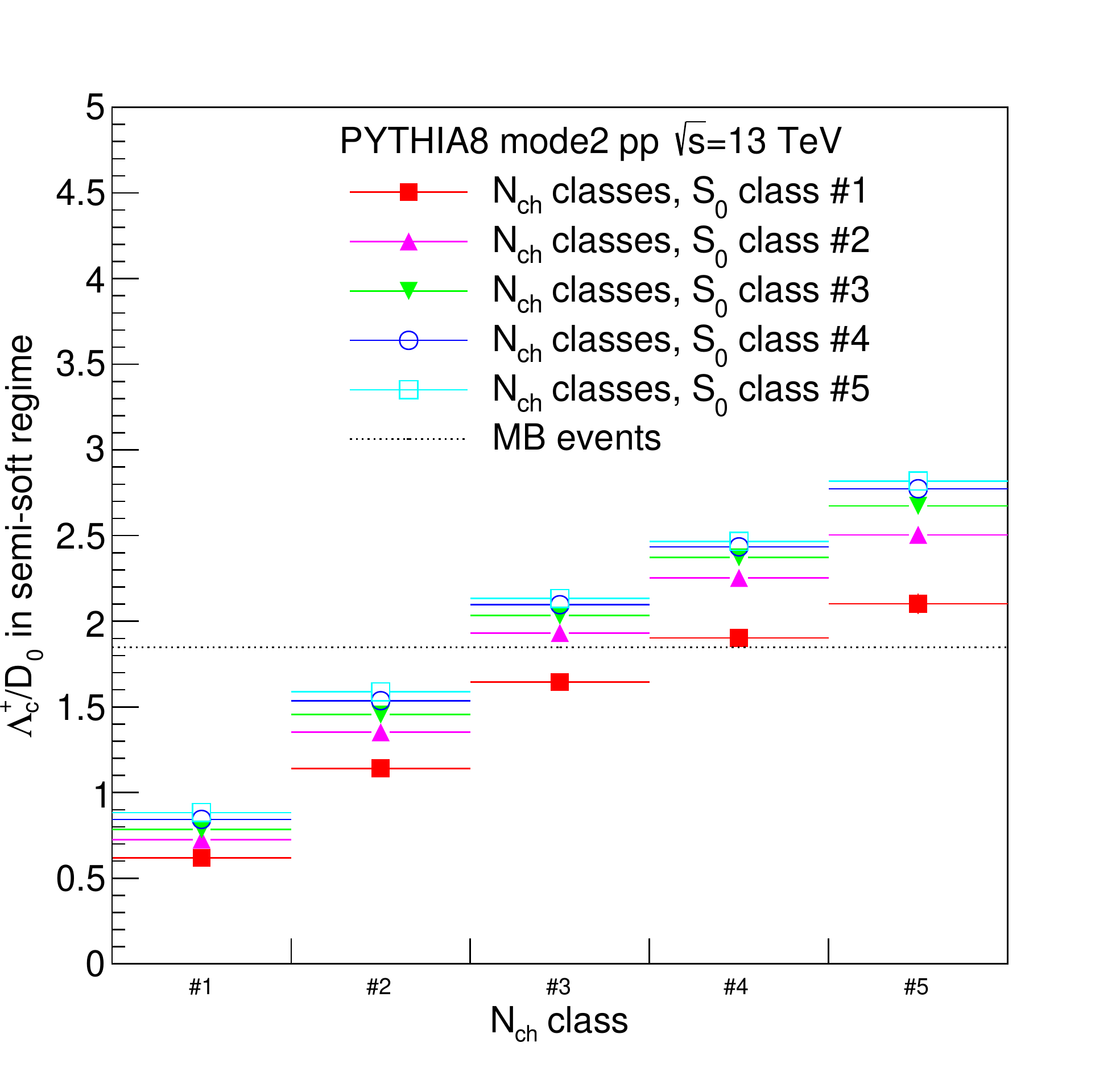}%
\caption{Left: \Lc to \Dz ratios in function of \pT, for different transverse spherocity bins for $\Nch>50$. Right: \Lc to \Dz ratios integrated over the coalescence regime $2<\pT<6$ \GeVc, in function of \Nch, in several spherocity bins. The dashed line represents the average of minimum-bias events.}
\label{fig:Sphero}
\end{figure}

The presence of a jet correlates with the event multiplicity. To minimize the effect of this correlation we observe spherocity in fixed multiplicity intervals. 
In Fig.~\ref{fig:Sphero} (left) we show the \LcToDz ratio for different spherocity bins for events with $\Nch > 50$. 
Jetty events show a significantly smaller \Lc enhancement than more isotropic events.
In the right panel we show the \LcToDz ratios for different \SO bins, integrated over the coalescence regime and plotted in function of the \Nch classes. 
It can be observed that for lower multiplicity events the dependence of the ratio on spherocity is weak, while for higher multiplicities the separation between the ratios for low and high \SO values becomes more significant. We can understand this by considering that in case of lower-multiplicity events most of the contribution comes from the leading process, without a significant contribution from the UE. For higher multiplicities, however, there is sufficient room for the underlying event. In the latter case both the leading process and the UE contribute to the event, and \SO characterizes their relative proportion.

\section{Conclusions}
\label{sec:conclusions}

Recent observations of low-\pT enhancement of charmed-baryon production relative to charmed mesons in proton--proton collisions at LHC energies questioned the universality of charm fragmentation and provide a challenge for theoretical models. 
In this work we demonstrated that appropriately defined event-activity classifiers provide great sensitivity to the production mechanisms of the \Lc baryon in proton--proton collisions at LHC energies.
Utilizing PYTHIA 8 simulations with enhanced color-reconnection we found that \LcToDz in hadron-triggered events shows a pronounced dependence on the transverse-event-activity classifier \RT, but not on the jet-activity classifier \RNC. We also showed that \LcToDz depends on the transverse spherocity $S_0$ in the case of events with high final-state multiplicity.
This shows that in the given scenario the excess \Lc production is primarily linked to the underlying event activity and not to the production of jets. Upcoming LHC data taking periods in Run-3 and beyond will allow for differential measurements of charmed-hadron production with an unprecedented precision. The comparison of \LcToDz ratios in hadron-triggered data in function of \RT and \RNC, as well as the double-differential evaluation of minimum-bias data in function of $S_0$ and \Nch opens up the possibility to differentiate between competing scenarios that describe flavor-dependent hadroproduction in the underlying event and within jets. 

\begin{acknowledgments}
The authors would like to thank for the many useful discussions they had with Antonio Ortiz and Fabio Colamaria. This work was supported by the Hungarian National Research, Development and Innovation Office (NKFIH) under the contract numbers OTKA FK131979 and K135515, and the NKFIH grant 2019-2.1.11-T\'ET-2019-00078. 
The authors acknowledge the computational resources provided by the Wigner GPU Laboratory and research infrastructure provided by the E\"otv\"os Lor\'and Research Network (ELKH).
\end{acknowledgments}

\nocite{*}
\bibliographystyle{utphys}
\bibliography{LcDbiblio}

\providecommand{\href}[2]{#2}\begingroup\raggedright\begin{thebibliography}{10}

\bibitem{Collins:1989gx}
J.~C. Collins, D.~E. Soper, and G.~F. Sterman, ``{Factorization of Hard
  Processes in QCD},'' \href{http://dx.doi.org/10.1142/9789814503266_0001}{{\em
  Adv. Ser. Direct. High Energy Phys.} {\bfseries 5} (1989) 1--91},
  \href{http://arxiv.org/abs/hep-ph/0409313}{{\ttfamily arXiv:hep-ph/0409313}}.

\bibitem{ALICE:2017thy}
{\bfseries ALICE} Collaboration, S.~Acharya {\em et~al.}, ``{$\Lambda_{\rm
  c}^+$ production in pp collisions at $\sqrt{s} = 7$ TeV and in p-Pb
  collisions at $\sqrt{s_{\rm NN}} = 5.02$ TeV},''
  \href{http://dx.doi.org/10.1007/JHEP04(2018)108}{{\em JHEP} {\bfseries 04}
  (2018) 108}, \href{http://arxiv.org/abs/1712.09581}{{\ttfamily
  arXiv:1712.09581 [nucl-ex]}}.

\bibitem{CMS:2019uws}
{\bfseries CMS} Collaboration, A.~M. Sirunyan {\em et~al.}, ``{Production of
  $\Lambda_\mathrm{c}^+$ baryons in proton-proton and lead-lead collisions at
  $\sqrt{s_\mathrm{NN}}=$ 5.02 TeV},''
  \href{http://dx.doi.org/10.1016/j.physletb.2020.135328}{{\em Phys. Lett. B}
  {\bfseries 803} (2020) 135328},
  \href{http://arxiv.org/abs/1906.03322}{{\ttfamily arXiv:1906.03322
  [hep-ex]}}.

\bibitem{ALICE:2020wfu}
{\bfseries ALICE} Collaboration, S.~Acharya {\em et~al.}, ``{$\Lambda_{\rm
  c}^{+}$ production and baryon-to-meson ratios in pp and p-Pb collisions at
  $\sqrt{s_\mathrm{NN}} = 5.02$ TeV at the LHC},''
  \href{http://arxiv.org/abs/2011.06078}{{\ttfamily arXiv:2011.06078
  [nucl-ex]}}.

\bibitem{Christiansen:2015yqa}
J.~R. Christiansen and P.~Z. Skands, ``{String Formation Beyond Leading
  Colour},'' \href{http://dx.doi.org/10.1007/JHEP08(2015)003}{{\em JHEP}
  {\bfseries 08} (2015) 003}, \href{http://arxiv.org/abs/1505.01681}{{\ttfamily
  arXiv:1505.01681 [hep-ph]}}.

\bibitem{Song:2018tpv}
J.~Song, H.-h. Li, and F.-l. Shao, ``{New feature of low $p_{T}$ charm quark
  hadronization in $pp$ collisions at $\sqrt{s}=7$ TeV},''
  \href{http://dx.doi.org/10.1140/epjc/s10052-018-5817-x}{{\em Eur. Phys. J. C}
  {\bfseries 78} no.~4, (2018) 344},
  \href{http://arxiv.org/abs/1801.09402}{{\ttfamily arXiv:1801.09402
  [hep-ph]}}.

\bibitem{Plumari:2017ntm}
S.~Plumari, V.~Minissale, S.~K. Das, G.~Coci, and V.~Greco, ``{Charmed Hadrons
  from Coalescence plus Fragmentation in relativistic nucleus-nucleus
  collisions at RHIC and LHC},''
  \href{http://dx.doi.org/10.1140/epjc/s10052-018-5828-7}{{\em Eur. Phys. J. C}
  {\bfseries 78} no.~4, (2018) 348},
  \href{http://arxiv.org/abs/1712.00730}{{\ttfamily arXiv:1712.00730
  [hep-ph]}}.

\bibitem{He:2019tik}
M.~He and R.~Rapp, ``{Charm-Baryon Production in Proton-Proton Collisions},''
  \href{http://dx.doi.org/10.1016/j.physletb.2019.06.004}{{\em Phys. Lett. B}
  {\bfseries 795} (2019) 117--121},
  \href{http://arxiv.org/abs/1902.08889}{{\ttfamily arXiv:1902.08889
  [nucl-th]}}.

\bibitem{ALICE:2021bli}
{\bfseries ALICE} Collaboration, S.~Acharya {\em et~al.}, ``{Measurement of the
  cross sections of $\Xi^0_{\rm c}$ and $\Xi^+_{\rm c}$ baryons and
  branching-fraction ratio BR($\Xi^0_{\rm c} \rightarrow \Xi^-{\rm e}^+\nu_{\rm
  e}$)/BR($\Xi^0_{\rm c} \rightarrow \Xi^-\pi^+$) in pp collisions at 13
  TeV},'' \href{http://arxiv.org/abs/2105.05187}{{\ttfamily arXiv:2105.05187
  [nucl-ex]}}.

\bibitem{Hills:2021eto}
{\bfseries ALICE} Collaboration, C.~Hills, ``{Charmed-baryon production and
  hadronization studies with ALICE},''
  \href{http://dx.doi.org/10.22323/1.387.0079}{{\em PoS} {\bfseries
  HardProbes2020} (2021) 079}.

\bibitem{Martin:2016igp}
T.~Martin, P.~Skands, and S.~Farrington, ``{Probing Collective Effects in
  Hadronisation with the Extremes of the Underlying Event},''
  \href{http://dx.doi.org/10.1140/epjc/s10052-016-4135-4}{{\em Eur. Phys. J. C}
  {\bfseries 76} no.~5, (2016) 299},
  \href{http://arxiv.org/abs/1603.05298}{{\ttfamily arXiv:1603.05298
  [hep-ph]}}.

\bibitem{Sjostrand:2014zea}
T.~Sj\"ostrand, S.~Ask, J.~R. Christiansen, R.~Corke, N.~Desai, P.~Ilten,
  S.~Mrenna, S.~Prestel, C.~O. Rasmussen, and P.~Z. Skands, ``{An introduction
  to PYTHIA 8.2}'' \href{http://dx.doi.org/10.1016/j.cpc.2015.01.024}{{\em
  Comput. Phys. Commun.} {\bfseries 191} (2015) 159--177},
  \href{http://arxiv.org/abs/1410.3012}{{\ttfamily arXiv:1410.3012 [hep-ph]}}.

\bibitem{Ortiz:2015ttf}
A.~Ortiz, G.~Pai\'c, and E.~Cuautle, ``{Mid-rapidity charged hadron transverse
  spherocity in pp collisions simulated with Pythia},''
  \href{http://dx.doi.org/10.1016/j.nuclphysa.2015.05.010}{{\em Nucl. Phys. A}
  {\bfseries 941} (2015) 78--86},
  \href{http://arxiv.org/abs/1503.03129}{{\ttfamily arXiv:1503.03129
  [hep-ph]}}.

\end{thebibliography}\endgroup

\end{document}